\documentclass[]{spie}  

\usepackage{amsmath,amsfonts,amssymb}
\usepackage{graphicx}
\usepackage{gensymb}
\usepackage[colorlinks=true, allcolors=blue]{hyperref}

\newcommand{\aap}{{\it Astron. Astrophys.}}

\newcommand{\procspie}{{\it Proc. SPIE}}

\newcommand{\nat}{{\it Nature}}

\title{The STRIP instrument of the Large Scale Polarization Explorer: microwave eyes to map the Galactic polarized foregrounds} 

\author[a,b]{Cristian~Franceschet}
\author[a,b]{Sabrina~Realini}
\author[a,b]{Aniello~Mennella}
\author[e]{Giuseppe~Addamo}
\author[c]{Alessandro~Ba\`u}
\author[p]{Paola~M.~Battaglia}
\author[a,b]{Marco~Bersanelli}
\author[a,b]{Barbara~Caccianiga}
\author[a,b]{Silvia~Caprioli}
\author[a,b]{Francesco~Cavaliere}
\author[k]{Kieran~A.~Cleary}
\author[d]{Francesco~Cuttaia}
\author[q]{Francesco~Del~Torto}
\author[g]{Viviana~Fafone}
\author[e]{Zunnoorain~Farooqui}
\author[i,j]{Ricardo~T.~G\'enova Santos}
\author[l]{Todd~C.~Gaier}
\author[c]{Massimo~Gervasi}
\author[c]{Tommaso~Ghigna}
\author[a,b]{Federico~Incardona}
\author[a]{Simone~Iovenitti}
\author[h]{Mike~Jones}
\author[k]{Pekka~Kangaslahti}
\author[c]{Roberto~Mainini}
\author[a,b]{Davide~Maino}
\author[f]{Michele~Maris}
\author[n]{Patricio~Mena}
\author[o]{Roc\'io~Molina}
\author[d]{Gianluca~Morgante}
\author[c]{Andrea~Passerini}
\author[i,j]{Maria~del~Rosario~Perez-de-Taoro}
\author[e]{Oscar~A.~Peverini}
\author[a]{Federico~Pezzotta}
\author[c]{Claudio~Pincella}
\author[n]{Nicol\'as~Reyes}
\author[g]{Alessio~Rocchi}
\author[i]{Jos\'e~A.~Rubi\~no--Mart\'in}
\author[d]{Maura~Sandri}
\author[f]{Stefano~Sartor}
\author[k]{Mary~Soria}
\author[o]{Valeria~Tapia}
\author[d]{Luca~Terenzi}
\author[a,b]{Maurizio~Tomasi}
\author[m]{Elisabetta~Tommasi}
\author[a]{Daniele~M.~Vigan\'o}
\author[d]{Fabrizio~Villa}
\author[e]{Giuseppe~Virone}
\author[m]{Angela~Volpe}
\author[h]{Bob~Watkins}
\author[f]{Andrea~Zacchei}
\author[c]{Mario~Zannoni}

\affil[a]{Physics Dept., Universit\`a degli Studi di Milano, Via Celoria 16, I-20133 Milano, Italy}
\affil[b]{INFN Sezione di Milano, Via Celoria 16, I-20133 Milano, Italy}
\affil[c]{Physics Dept., University of Milano--Bicocca, Piazza della Scienza 3, I-20126 Milano, Italy}
\affil[d]{INAF - Osservatorio di Astrofisica e Scienza dello Spazio di Bologna, Via Gobetti 101, 40129 Bologna, Italy}
\affil[e]{IEIIT-CNR, Politecnico di Torino, Corso Duca degli Abruzzi 24, 10129 Torino, Italy}
\affil[f]{INAF - Osservatorio Astronomico di Trieste, Via G.B. Tiepolo 11, 34143 Trieste, Italy}
\affil[g]{INFN Roma Tor Vergata, 00133 Roma, Italy}
\affil[h]{Department of Physics (Astrophysics), University of Oxford, Oxford OX1 3RH, U.K.}
\affil[i]{Instituto de Astrof\'{\i}sica de Canarias (IAC), E-38200 La Laguna, Tenerife, Spain}
\affil[j]{Departamento de Astrof\'{\i}sica, Universidad de La Laguna, E-38206 La Laguna, Tenerife, Spain}
\affil[k]{Department of Astronomy, California Institute of Technology, 1200 East California Boulevard, Pasadena, CA 91125, USA}
\affil[l]{Jet Propulsion Laboratory, California Institute of Technology, 4800 Oak Grove Drive, Pasadena, CA 91109, USA}
\affil[m]{Agenzia Spaziale Italiana, Via del Politecnico snc, 00133 Roma, Italy}
\affil[n]{Electrical Engineering Department, University of Chile, Av. Tupper 2007, Santiago, Chile}
\affil[o]{Astronomy Department, University of Chile, Camino El Observatorio 1515, Santiago, Chile}
\affil[p]{IASF-Mi, Istituto Nazionale di Astrofisica, Via Bassini 15, Milano, Italy}
\affil[q]{Altran Italia S.p.A., Corso Sempione 66, 20154 Milan, Italy}

\authorinfo{Further author information: (Send correspondence to C. Franceschet)\\C. Franceschet: E-mail: cristian.franceschet@fisica.unimi.it}

\begin{document} 
\maketitle

\begin{abstract}
In this paper we discuss the latest developments of the STRIP instrument of the ``Large Scale Polarization Explorer'' (LSPE) experiment. LSPE is a novel project that combines ground-based (STRIP) and balloon-borne (SWIPE) polarization measurements of the microwave sky on large angular scales to attempt a detection of the ``B-modes'' of the Cosmic Microwave Background polarization. STRIP will observe approximately 25\% of the Northern sky from the ``Observatorio del Teide'' in Tenerife, using an array of forty-nine coherent polarimeters at 43~GHz, coupled to a 1.5~m fully rotating crossed-Dragone telescope. A second frequency channel with six-elements at 95~GHz will be exploited as an atmospheric monitor. At present, most of the hardware of the STRIP instrument has been developed and tested at sub-system level. System-level characterization, starting in July 2018, will lead STRIP to be shipped and installed at the observation site within the end of the year. The on-site verification and calibration of the whole instrument will prepare STRIP for a 2-years campaign for the observation of the CMB polarization.

\end{abstract}


\section{INTRODUCTION}
\label{sec:intro}  

The measurement of the polarized component of the Cosmic Microwave Background (CMB) represents the new frontier of the observational cosmology. CMB polarization encodes information complementary to that extracted from temperature anisotropies alone, breaking the degeneracies among some cosmological parameters, while shedding light on the very first moments of our Universe. The distribution of CMB polarization in the sky is usually decomposed into a gradient and a curl component, called E-modes and B-modes, respectively. E-modes, whose signal is at the level of the $\mu$K, have been widely observed since their first detection in 2002 by DASI\cite{DASI}. In 2015, Planck provided a full-sky map of the E-modes of the CMB polarization\cite{Planck_comp_sep} and of its astrophysical contaminants\cite{Planck_foreground} in a frequency range from 30 to 353 GHz.

On the contrary, no experiment has detected primordial B-modes, yet. A detection of primordial B-modes would provide a strong evidence in favor of the inflationary paradigm. In fact, according to inflation, primordial tensor perturbation (as opposed to density or scalar perturbations) resulted in a stochastic background of gravitational waves, which would have left a distinctive imprint on the polarization pattern of CMB photons at the recombination (and at the later epoch of the reionization). The amplitude of the B-mode component is parametrized by the tensor-to-scalar ratio $r$ and, to date, only upper bounds have been set by measurements. The joint BICEP2/Keck Array and Planck analysis\cite{joint_analysis} indicates $r < 0.07$ at 95\% C.L., showing strong evidence for dust and no statistically significant evidence for tensor modes.

Besides a strict control of the instrumental noise and systematic effects, a deep knowledge of astrophysical foregrounds is a key element for polarization analysis. In fact, both synchrotron and thermal dust emissions are partially linearly polarized, with a variable polarization level from point to point in the sky.

In the recent study reported in Ref.~\citenum{spass_synch}, the S-PASS maps of the Southern Sky emission in linear polarization at 2.3 GHz are used to assess the polarized synchrotron contamination to CMB observations of the B-modes at frequencies up to 150 GHz. The study points out that, at 90 GHz, the minimal contamination in the cleanest regions of the observed sky is at the level of equivalent tensor-to-scalar ratio $r_{synch}\approx 10^{-3}$, as shown in Fig. 11 of Ref.~\citenum{spass_synch}. By combining S-PASS data with Planck 353 GHz observations, the map of the minimum level of total polarized foreground contamination to B-modes shows that there is no region of the sky (within the sky portion covered by the S-PASS survey), at any frequency, where Galactic foreground contamination lies below a B-mode signal with $r\approx 10^{-3}$. Therefore, frequency channels monitoring foreground emissions, on both low and high frequency, are mandatory for all future experiments aiming at the observation of the primordial gravitational waves imprint on the CMB.
   
The quest for the B-modes is carried out by several CMB polarization experiments, from either ground or balloon. Furthermore, the scientific community is actively working on proposals for a future space mission dedicated to CMB polarization. In this context, we discuss the STRIP instrument of the ``Large Scale Polarization Explorer'', the Italian Space Agency’s upcoming experiment for the observation of the CMB polarization on large angular scales.

\section{The ``Large Scale Polarization Explorer''}
The Large Scale Polarization Explorer (LSPE) combines ground-based and balloon-borne observations and aims at improving the limit on the tensor-to-scalar ratio down to $r$ = 0.03 at 99.7\% confidence level. A second target is to produce wide maps of foreground polarization generated in our Galaxy by synchrotron and interstellar dust emissions in a range of frequencies between 40 and 250 GHz. These will be important to map Galactic magnetic fields and to study the properties of ionized gas and of diffuse interstellar dust in our Galaxy\cite{LSPE}.

LSPE will observe a 25\% fraction of the sky in the Northern hemisphere by means of two instruments, complementary for frequency coverage and technology, SWIPE and STRIP. SWIPE\cite{SWIPE} consists of three arrays of 110 large throughput multi-mode bolometers centered at the frequencies 140, 220 and 240 GHz. 
A rotating Half Wave Plate polarization modulator is employed to mitigate the systematic effects due to instrumental non-idealities. SWIPE will survey the Northern Sky from a spinning stratospheric balloon, launched from the Svalbard Islands, in a long duration flight (approximately 14 days) during the Arctic winter.

Initially designed to share the same gondola of SWIPE\cite{STRIP}, STRIP has been recently reconverted into a ground-based instrument, while providing approximately the same sky coverage (observed sky overlap $>80\%$), so that LSPE is still conceived as a single experiment. STRIP consists of an array of 49 polarimeters in the Q-band (43~GHz) and a 6-elements W-band (95~GHz) atmospheric monitor. Receivers are cooled down to 20~K in a cryostat facing the focal surface of a dual reflector telescope, with a 1.5~m primary mirror. 
STRIP will start operations by the end of 2018 at the ``Observatorio del Teide'' in Tenerife, sharing the observing site with the QUIJOTE experiment\cite{QUIJOTE} which currently consists of two telescopes in the 10-20 GHz and 30 GHz frequency bands. The cooperation established between the two experiments teams, together with the availability of full sky measurements provided by Planck and WMAP, will add further strength to the characterization of the foreground with sensitivities at the state-of-the-art and in a wide range of frequencies.

In the following, we discuss the status and the latest developments of the STRIP instrument and provide a brief overview its main requirements focusing on sensitivity and scanning strategy.

\section{STRIP SCIENTIFIC REQUIREMENTS} 

Table~\ref{tab:requirements} summarizes the STRIP main scientific requirements for the two channels in the Q- and W-band. Sensitivity is given in thermodynamic temperature units and assumes 24 months of observations with a conservative 35\% duty cycle (the fraction of observation time that will be useful from the scientific point of view). The Q-band goal sensitivity corresponds to an improvement of a factor 5 over the Planck-LFI 44 GHz sensitivity\cite{Planck_sensitivity} on the same pixel size at the end of the 30 months mission.

\begin{table}[h]
\caption{STRIP main scientific requirements. }
\label{tab:requirements}
\begin{center}       
\begin{tabular}{|l|c|c|} 
\hline
\rule[-1ex]{0pt}{3.5ex} & {\bf Q-band} & {\bf W-band}\\
\hline
\rule[-1ex]{0pt}{3.5ex} Angular resolution ($\degree$) & \multicolumn{2}{|c|}{at least 0.5$\degree$}\\
\hline
\rule[-1ex]{0pt}{3.5ex} Observed sky fraction (\%) & \multicolumn{2}{|c|}{$\geq 18$}\\
\hline
\rule[-1ex]{0pt}{3.5ex} Sensitivity per $1\degree$ pixel $\delta Q(U)_{deg} $ ($\mu$K) & $\le 1.2$ & $\le 4.5$\\
\hline
\end{tabular}
\end{center}
\end{table} 

The STRIP Q-band sensitivity on Stokes parameters $Q$ and $U$, $\delta Q(U)_{deg} \le 1.2$ $\mu$K can be obtained in a 25\% fraction of the sky assuming an atmospheric emission of 9~K and an instrument system temperature better than $\sim$35~K with a 18\% bandwidth. These numbers include, in addition to polarimeter noise, extra contributions from the telescope, the cryostat window, the feed system and back-end electronics. 


The STRIP focal plane will also host a 6-elements array in the W-band for monitoring the atmospheric contribution. Characterizing the Tenerife observing site at W-band, i.e., near the foregrounds minimum, is of high interest not only for LSPE, but also for potential future CMB experiments covering the Northern hemisphere.

By assuming the same observational parameters (25\% sky fraction and 24-months observation time with 35\% duty cycle), an instrument system temperature of approximately 62~K and an atmospheric contribution of 11~K, the final sensitivity per $1\degree$ pixel will be $\delta Q(U)_{deg} \le 4.5~\mu$K. Regarding the W-band channel, we do not have a CMB-driven requirement on sensitivity, but the signal-to-noise needs be as high as possible for monitoring the atmosphere at the Tenerife site and, possibly, detecting any polarized emission.


\section{Nominal scanning Strategy}
\label{sec:scan_strat}

As detailed in this proceedings, the STRIP nominal scanning strategy is driven by the following criteria: (1) matching a wide sky coverage with uniform sensitivity distribution, (2) minimizing the radiation pickup from
ground and (3) the availability of polarized sources during observation, such as planets, the Crab Nebula (for absolute calibration), the Moon (useful for polarization angle verification), Perseus molecular complex and Orion Nebula.

\begin{figure}
   \begin{center}
   \includegraphics[width=0.67\textwidth]{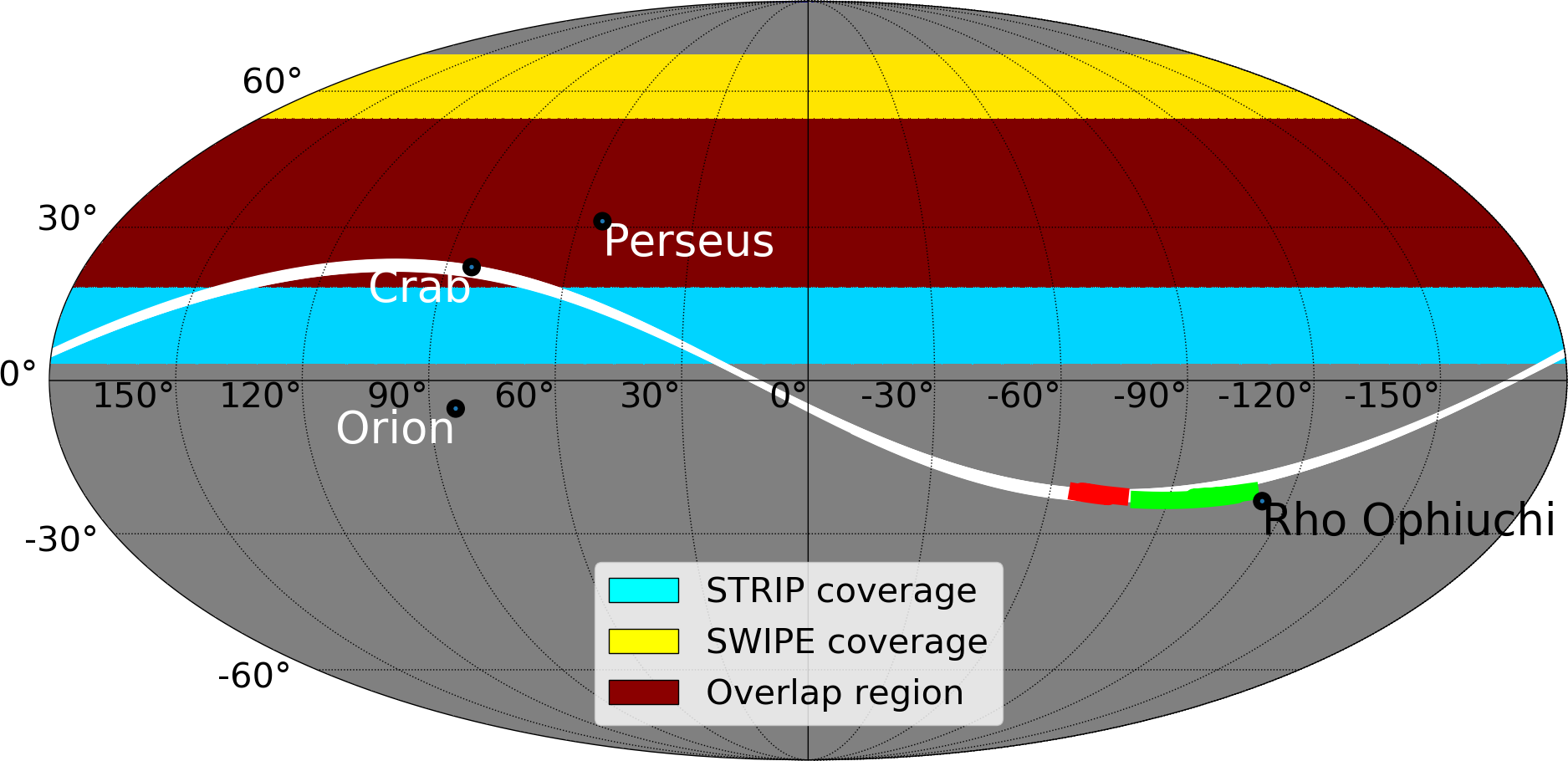}
   \end{center}
   \caption[sky_coverage] 
   { \label{fig:sky_coverage} 
Map in ICRS (equatorial) coordinates of the STRIP and SWIPE instruments sky coverage\cite{scanning_strategy}. The yellow area represents the SWIPE sky coverage, the cyan area represents the STRIP sky coverage, the dark-red area is the overlap. The map also shows the positions of the Crab and Orion nebulas, of Perseus molecular complex and the trajectories of Jupiter (black line), Saturn (red line) and the Moon (white line), as calculated for the year 2019.}
   \end{figure}

As a baseline, STRIP will scan the sky with a continuous azimuthal spin at 1 r.p.m., while maintaining telescope boresight at a fixed elevation angle (approximately 20 degrees from the Zenith at Tenerife latitude). By combining the telescope spinning with the daily rotation of the Earth around its axis, STRIP will cover a region of the Northern Sky that overlaps at least 80\% of the SWIPE survey, as shown in Fig.~\ref{fig:sky_coverage}. The resulting sky coverage of both instruments is $\sim$25\%.

\section{INSTRUMENT OVERVIEW AND STATUS}
In this section we outline the STRIP instrument design and provide an overview of its main sub-systems, i.e., optics, polarimeters and passive focal plane elements, cryogenic design.\cite{design_report}.

\subsection{The STRIP optical system}

\subsubsection{Optics}

The STRIP telescope is a Dragonian cross-fed dual-reflector system with a projected diameter aperture of 1.5~m. The Crossed-Dragone design gives exceptionally low aberrations and cross-polarization across a large focal plane, allowing us to feed a large number of detectors without requiring additional focusing optics that may introduce aberrations or cross-polarization. The Dragonian configuration is the best in terms of polarization purity and symmetry over a wide focal region\cite{tran}. 

The primary mirror of the STRIP telescope has an offset parabolic shape, with a 1500 mm aperture. The secondary mirror is an offset concave hyperboloid, with a 1719.76 mm $\times$ 1658.7 mm wide elliptical rim. The telescope provides an angular resolution of about $20'$ in Q-band.
The entire dual-reflector system has an equivalent focal length of 2700 mm, resulting in a F\#$\sim 1.8$.

\begin{figure}
   \begin{center}
   \begin{tabular}{c c}
   \includegraphics[width=0.49\textwidth]{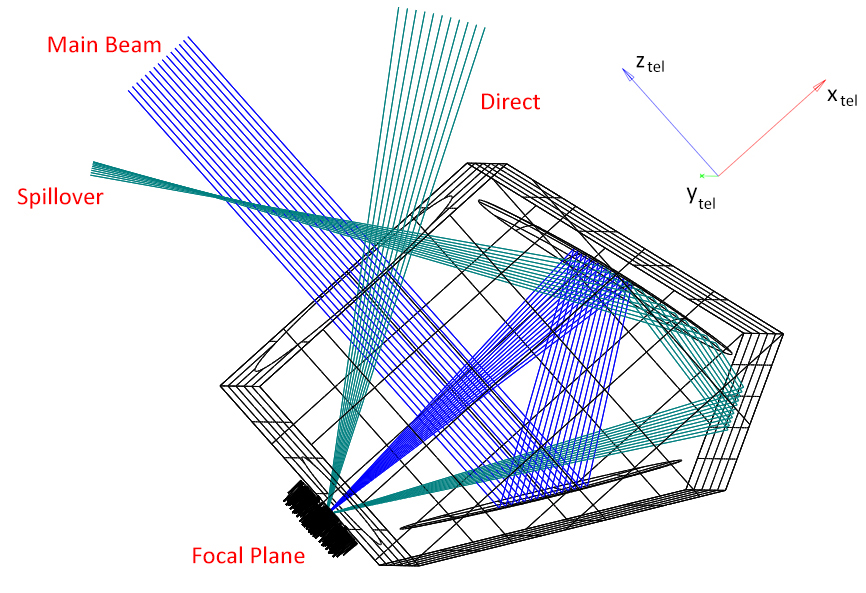}&
   \includegraphics[width=0.47\textwidth]{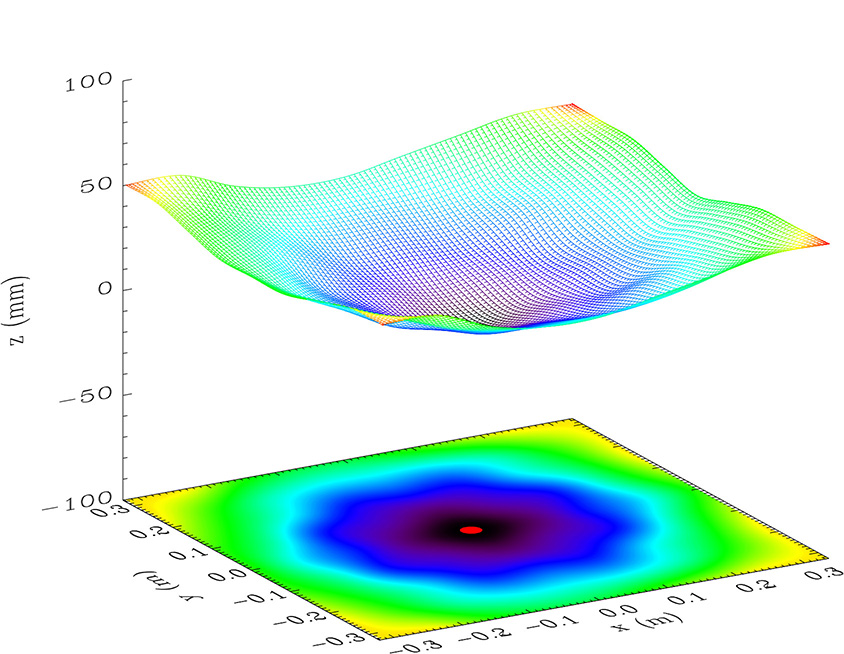}
   \end{tabular}
   \end{center}
   \caption[Telescope] 
   { \label{fig:telescope} 
{\it Left:} Telescope and shields geometry with the ray-tracing of some MrGTD contributions in the symmetry plane. {\it Right:} Focal surface of the STRIP telescope.}
   \end{figure}

The main requirements on the optics are a cross-polar discrimination better than -30 dB and a level of sidelobes rejection of -55 dB and -65 dB for near and far sidelobes, respectively. The whole array of feedhorns is placed in the focal region of the telescope, ensuring no obstruction of the field of view. 
For an optimal response in terms of resolution, directivity and sidelobes level, the feedhorn modules have been placed according to telescope focal surface shown in Fig.~\ref{fig:telescope}.

Originally designed and manufactured for the CLOVER experiment\cite{clover}, each of the mirrors has been machined from a single piece of aluminum. The back of the mirror is light-weighted by machining away most of the backing material, leaving some ribs to provide structural strength and location for three adjustable mounting points per mirror. 

We modeled STRIP optics with the software GRASP\footnote{The GRASP software is developed by TICRA (Copenhagen, DK) for analysing general reflector antennas (http://www.ticra.it).} using Physical Optics (PO) and Multi-Reflector Geometrical Theory of Diffraction (MrGTD). The model includes the two nominal reflectors, forty-nine Q band feedhorns and six W-band feedhorns, and the shielding structures, as shown in Fig.~\ref{fig:telescope}.

\begin{figure}
   \begin{center}
   \begin{tabular}{c c}
   \includegraphics[width=0.475\textwidth]{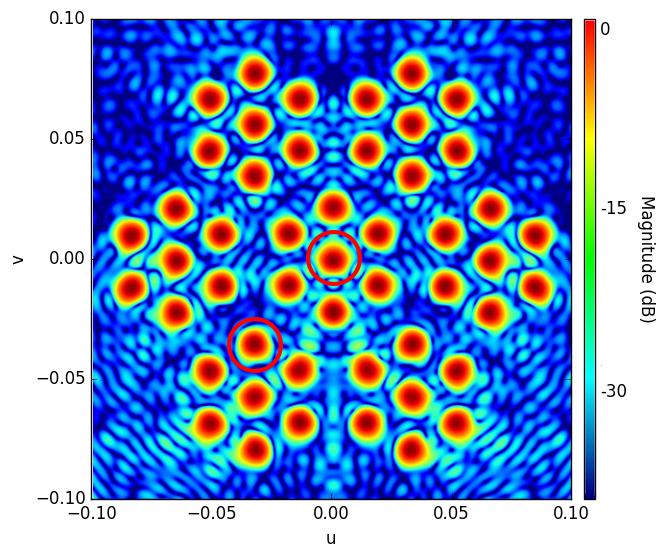}&
   \includegraphics[width=0.47\textwidth]{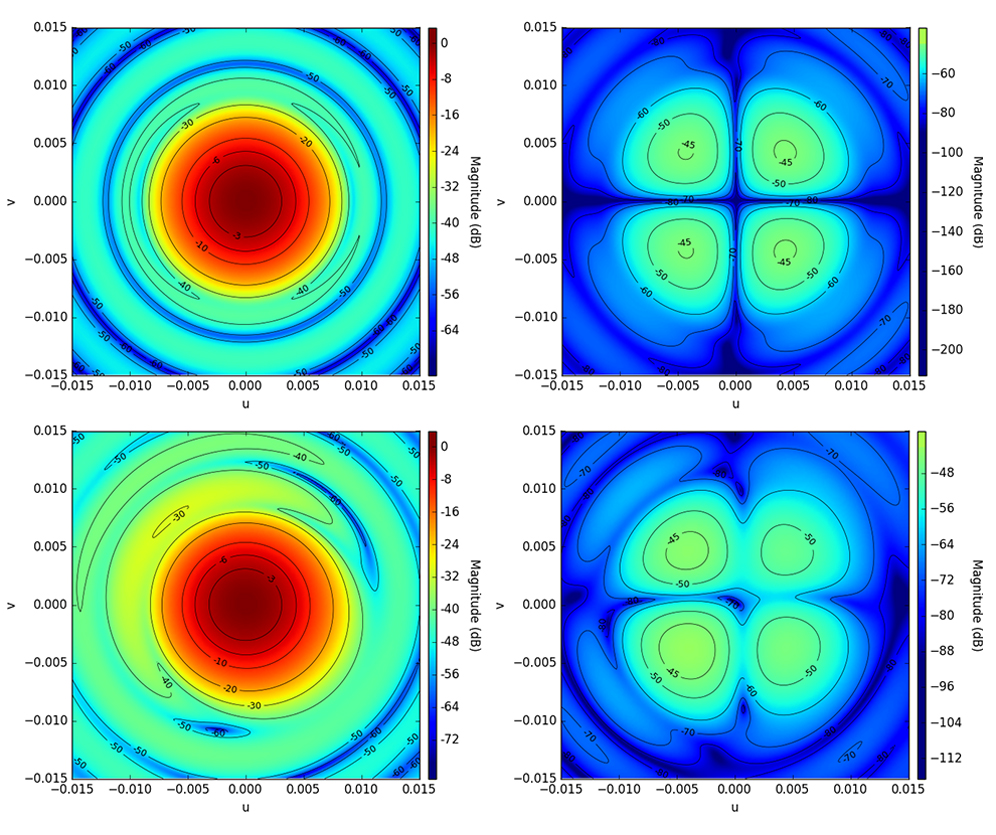}
   \end{tabular}
   \end{center}
   \caption[Footprint] 
   { \label{fig:footprint} 
{\it Left:} Footprint of the STRIP focal plane on the sky as seen by an observer looking towards the telescope along its optical axis. The origin of the uv-coordinate system is at the center of the focal plane. The z-axis is along the line-of-sight and points towards the observer. The red circles indicate the two beams represented in the right panel of this figure. {\it Right:} Contour plot in the uv-plane ($−0.015 < u, v < 0.015$) of the main beam co-polar (left side) and cross-polar (right side) component computed at 43 GHz for the feedhorn in the focus (first row) and for an off-axis feedhorn (second row), assuming an ideal telescope. The lines in the contour plots represent levels of at -3, -10, -20, -30, -40, -50, and -60 dB.}
   \end{figure}

The radiation pattern of the feedhorns in the STRIP optics have been simulated in their main beam coordinate system at 43 and 95 GHz, exciting each feedhorn individually with a linearly polarized signal. Fig.~\ref{fig:footprint} shows the footprint of the forty-nine beams at 43 GHz in the sky.
Main beams and sidelobes contributions have been simulated and analyzed separately, because of the different sampling in the $\theta$ and $\phi$ angular coordinates.

Given the telescope configuration and the feedhorn off-axis location on the focal surface, the main beams are not perfectly Gaussian, so that they cannot be described by a single parameter. For a complete characterization of the main beams, several descriptive parameters have been evaluated: the angular resolution (FWHM), the ellipticity (e), the main beam directivity (D), the cross-polar discrimination factor (XPD) and the spillover.
The angular resolution is around $21'$ and the directivity is around 54.7 dBi for all the Q-band channels, while the XPD ranges from 40.8 to 44.5 dB.

In order to reduce the contamination due to the sidelobes, the receiver and optics are surrounded by a co-moving baffle. Therefore, the electromagnetic model takes into account the shielding structures which redistribute the power that is radiated by the feedhorns and is not reflected by the telescope (see Fig.~\ref{fig:telescope}).

We analyzed each contribution to the sideobes using the MrGTD add-on of the GRASP software, which computes the scattered field from the reflectors performing a backward ray tracing. In principle, Physical Optics is the most accurate method for predicting beams and should be used in all regions surrounding the optical system. However, in the sidelobes region, the PO integrand oscillates and a finer integration grid is required, leading to an increasing computation time. For this reason the MrGTD represents a suitable method for predicting the full-sky radiation pattern of complex mm-wavelength optical systems.

\begin{figure}
   \begin{center}
   \begin{tabular}{c}
   \includegraphics[width=0.8\textwidth]{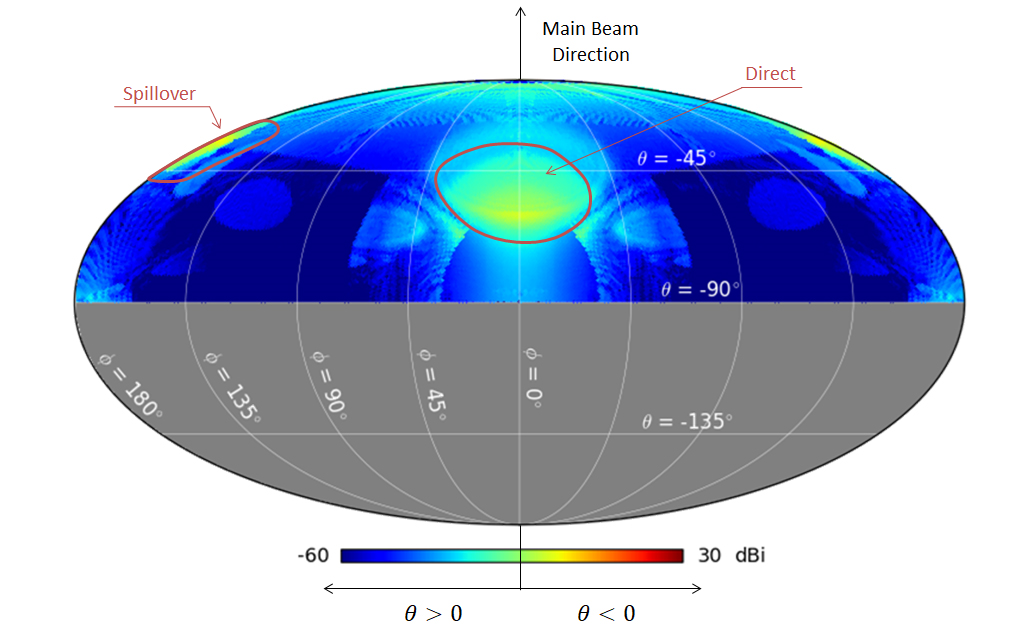}
   \end{tabular}
   \end{center}
   \caption[Mappa43] 
   { \label{fig:mappa} 
Far sidelobes at 43 GHz for the feedhorn placed in the telescope focus. The main beam points to the top of the map ($\theta = 0^{\circ}$) and peaks at 54.76 dBi. The direct contribution is due to the feedhorn sidelobe and peaks at about -5 dBi, while the spillover is due to a double reflection inside the shielding structure and peaks at about -0.8 dBi.}
   \end{figure}

We computed the sidelobes of the STRIP telescope up to the 2$^{nd}$ order of interaction (i.e. reflection or diffraction). They are unevenly distributed and concentrated mainly in two areas, namely the ``direct contribution'' and the ``spillover'' (see Fig.~\ref{fig:mappa}). The direct contribution is generated by the rays entering the feedhorns without any interaction with the reflectors; its shape and power level are given by the feedhorn radiation pattern pointing at about 60$^{\circ}$ from the line of sight of the telescope. The spillover is primarily due to rays reflected by one of the shields and then reflected by the sub-reflector. 

We are currently extending the above optical analysis to the W-band channel. We performed simulations of the W-band feedhorns response, as a reference for the on-going measurements in anechoic chamber. Furthermore, at present, GRASP simulations of the W-band main beams and sidelobes are running.

\subsubsection{Mount}

The telescope mount and optical assembly allow the STRIP optics and receiver to be pointed at any direction in the sky, with great flexibility for scanning strategy. The mount provides the mechanism for moving the optical assembly. The optical assembly holds the optics, receivers and some ancillary equipment, like the star tracker and the near-field calibrator.

\begin{figure}
   \begin{center}
   \begin{tabular}{c}
   \includegraphics[width=0.5\textwidth]{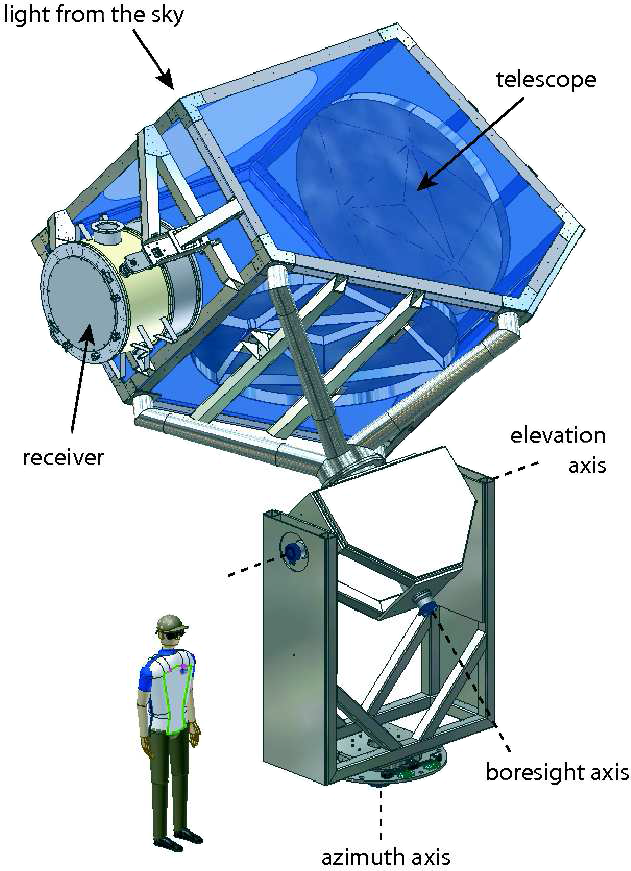}
   \end{tabular}
   \end{center}
   \caption[Mount] 
   { \label{fig:mount} 
Model of the three-axis mount which allows the rotation of the entire optical assembly around the telescope boresight. The mirrors are held inside a co-moving baffle (translucent in the figure)\cite{clover}.}
   \end{figure}

The telescope mount, already built for the CLOVER experiment, is a three axis system which can rotate fully in azimuth and reach elevations from 0$^{\circ}$ to 89.5$^{\circ}$. As a further modulation of the polarization signal, the mount is able to rotate the telescope around the pointing direction (see Fig.~\ref{fig:mount}).
The mount is able to carry out constant elevation scans at speeds up to 10$^{\circ}$/s with turnaround accelerations up to 20$^{\circ}$/s$^2$. 
Elastic deflection in the frame should be minimized, although recoverable with a pointing model. The current design provides a maximum deflection of about 0.1 mm which gives an accuracy slightly better than 20 arc seconds.
An integrated rotary joint will transmit fluid (air), power and data to the movable parts of the telescope and to the instrument.

The mount painting is a three coat process with a electrodeposited primer to provide a good chemical bond to steel, a Titanium dioxide based powder coat pigment layer and a tough clearcoat lacquer.
This paint will provide the best resistance to abrasion and a good corrosion protection. Eventually, the frames are painted white to minimize the solar heating of the structure.

The ambient temperatures at the observing site usually remain between 0 and 24 $^{\circ}$C. The average expected wind speed from any direction is 20 km/h, with occasionally wind gusts that can reach 200 km/h speed, in extreme weather conditions.
Under bad weather the mount will be protected by a sliding roof, which will avoid also exposure to wind-driven snow. This may occur in situations where it is not possible to access the site.
   

\subsection{The Focal Plane Unit}
The STRIP focal plane unit consists of an array of forty-nine coherent polarimeters operating in a 18\% frequency band centered at 43 GHz (Q-band) and six polarimeters operating in a frequency band centered at 95 GHz (W-band) for atmosphere monitoring. The receivers are based on the design developed for the QUIET ground-based experiment\cite{quiet_pol} and are able to simultaneously detect the Q and U Stokes parameters of the CMB.

The whole array of detectors will be placed in the focal region of the telescope, ensuring no obstruction of the field of view. All feedhorns modules have been placed on the telescope focal surface and oriented towards the main reflector center, so that an optimum spillover is obtained while guaranteeing low level of cross--polarization contamination.

\subsubsection{The Q-band Radiometer Chain}

\begin{figure}
   \begin{center}
   \begin{tabular}{c}
   \includegraphics[width=\textwidth]{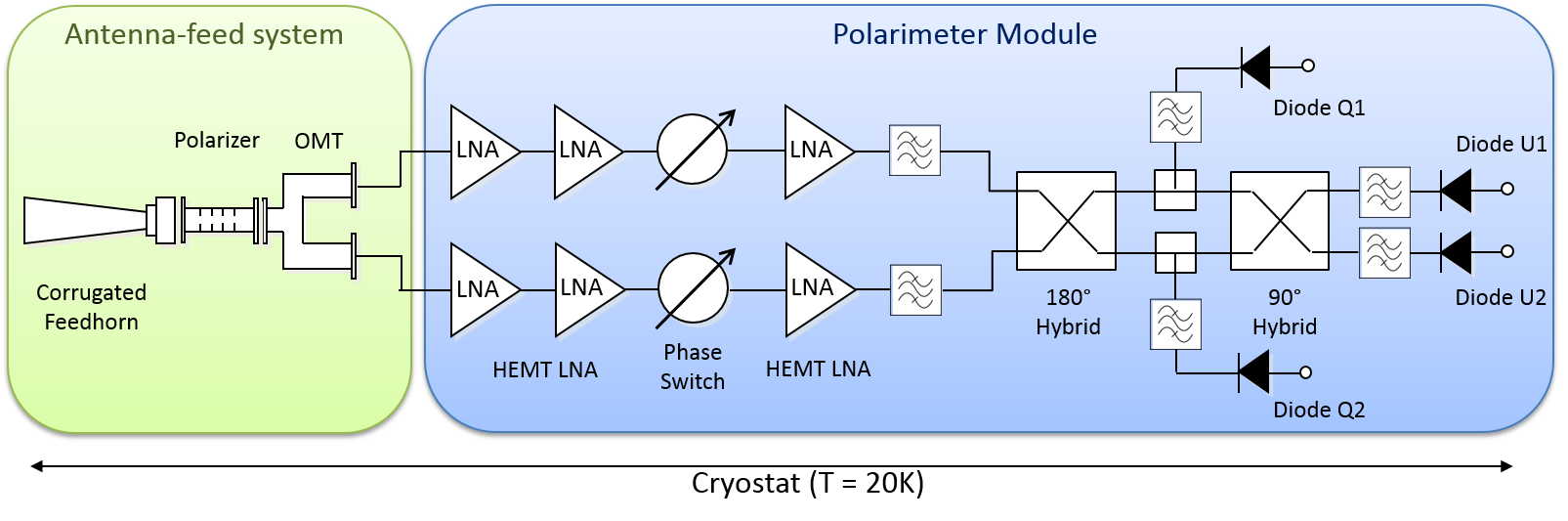}
   \end{tabular}
   \end{center}
   \caption[Configuration] 
   { \label{fig:config} 
Configuration of each Q-band receiver in the STRIP instrument. As explained in the text, the W-band receiver shares the same conceptual scheme, with the exception of the passive feeding network that implements the dual circular polarization by means of a septum polarizer, instead of the polarizer and OMT assembly.}
   \end{figure}

Each Q-band receiver includes a circular corrugated feedhorn, a polarizer, an orthomode transducer (OMT) and a polarimeter module working in the 39--48 GHz band. The forty-nine detectors are arranged into seven independent modules placed in the focal region of the dual-reflector telescope, and they are actively cooled down to 20 K. Figure~\ref{fig:config} shows a schematic of the Q-band receivers configuration.
   
The antenna at the beginning of each radiometric chain is a corrugated feedhorn, which has been chosen as the best solution in terms of beam symmetry, cross-polar response and sidelobes level.
The feedhorn design is dual-profiled with a \textit{sin\textsuperscript{2}} section for a length of 74 mm from the throat and an exponential section up to the aperture. The circular input waveguide diameter is 6.8~mm. This design is the best compromise between specifications and compactness.

\begin{figure}
   \begin{center}
   \begin{tabular}{c c}
   \includegraphics[width=0.4\textwidth]{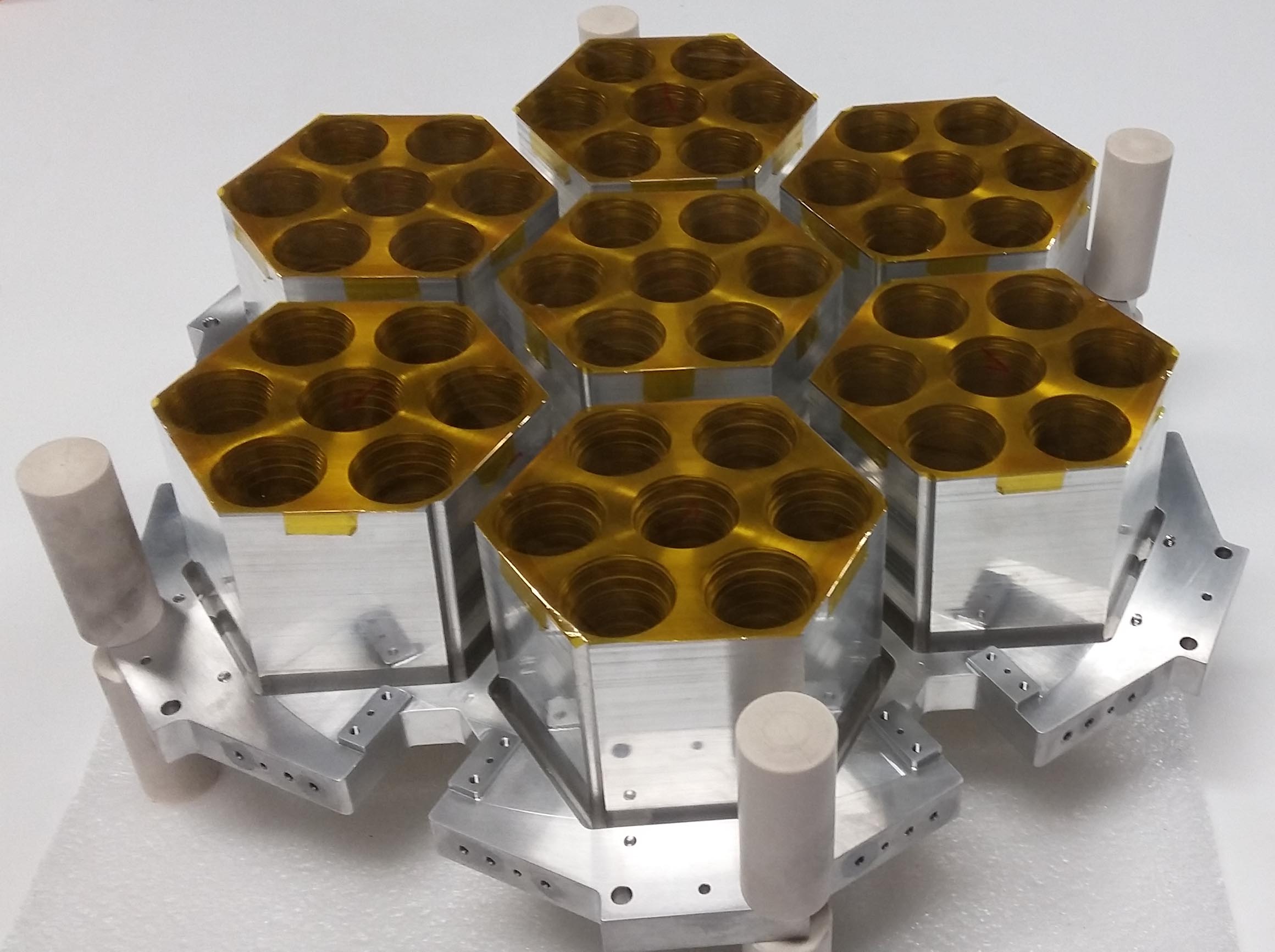} &
   \includegraphics[width=0.4\textwidth]{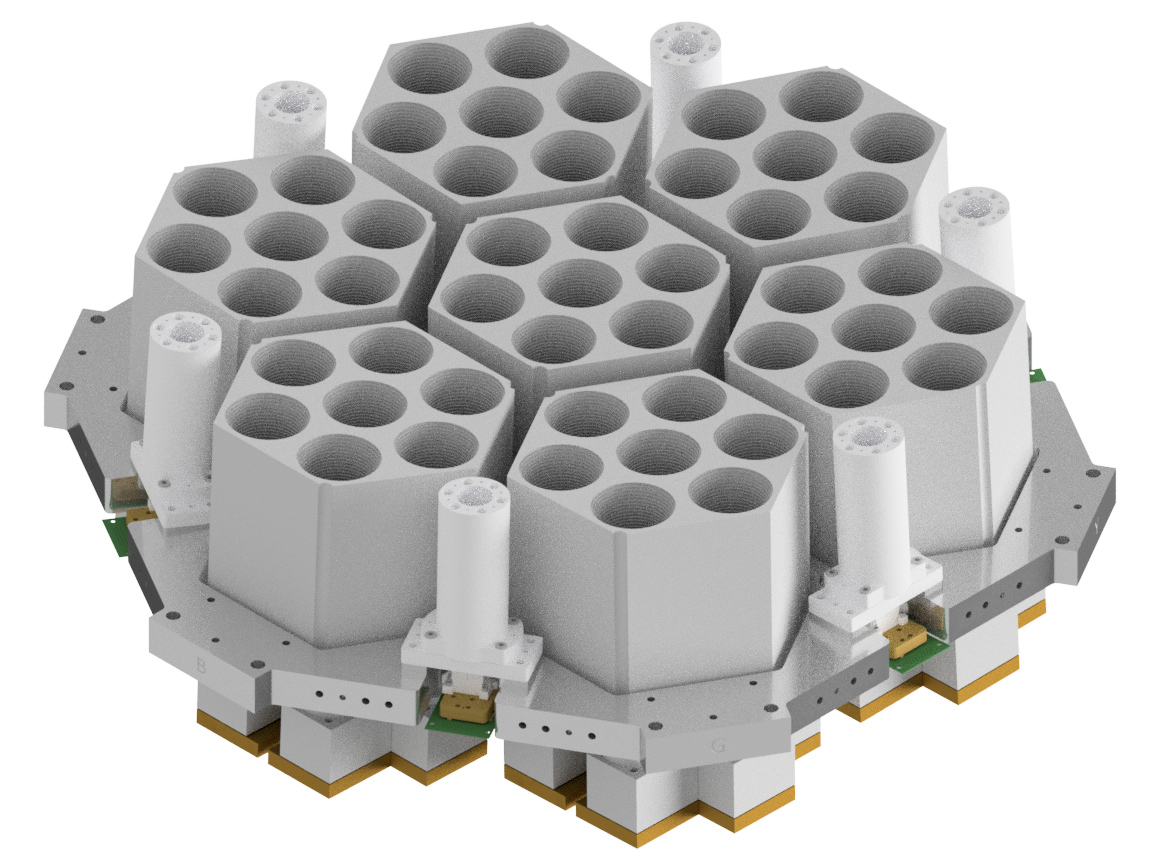}
   \end{tabular}
   \end{center}
   \caption[Frame] 
   { \label{fig:frame} 
{\it Left}: Picture of the seven modules assembled into the focal plane mechanical structure. {\it Right}: A rendered view of the fully integrated focal plane unit, including Q- and W-band radiometer chains.}
   \end{figure}

The 49 feedhorns of the array have the same profile and they are arranged in a honeycomb lattice of seven hexagonal modules, each including seven feedhorns (see Fig.~\ref{fig:frame}). Each 7-elements module is built with the platelet technique\cite{platelet_deltorto,feed_system_lspe,7x7_deltorto}, which consists in constructing the mechanical profile by stacking up Aluminum plates. Each plate is worked to obtain a tooth and a grove of a corrugation of the feed. Besides drilling holes in the platelets to form the feedhorns, additional cavities have been created to lighten the structure.
The mechanical tolerance of each plate has been verified with a metrological instrument, resulting in a maximum discrepancy of order 0.02-0.03 mm from the nominal dimensions (errors $< 0.5\%$ of a wavelength).

\begin{figure}
   \begin{center}
   \begin{tabular}{c c}
   \includegraphics[width=0.45\textwidth]{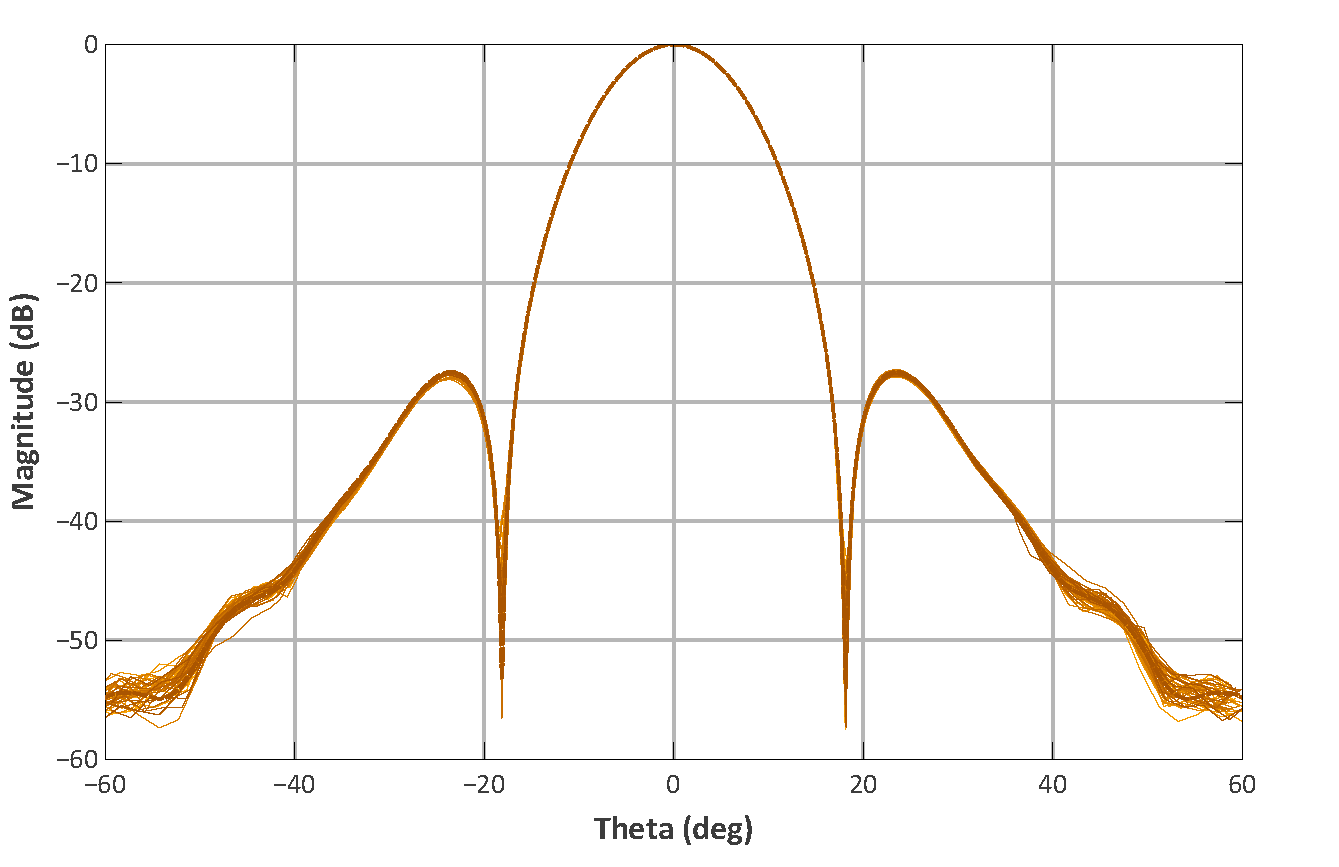} &
   \includegraphics[width=0.45\textwidth]{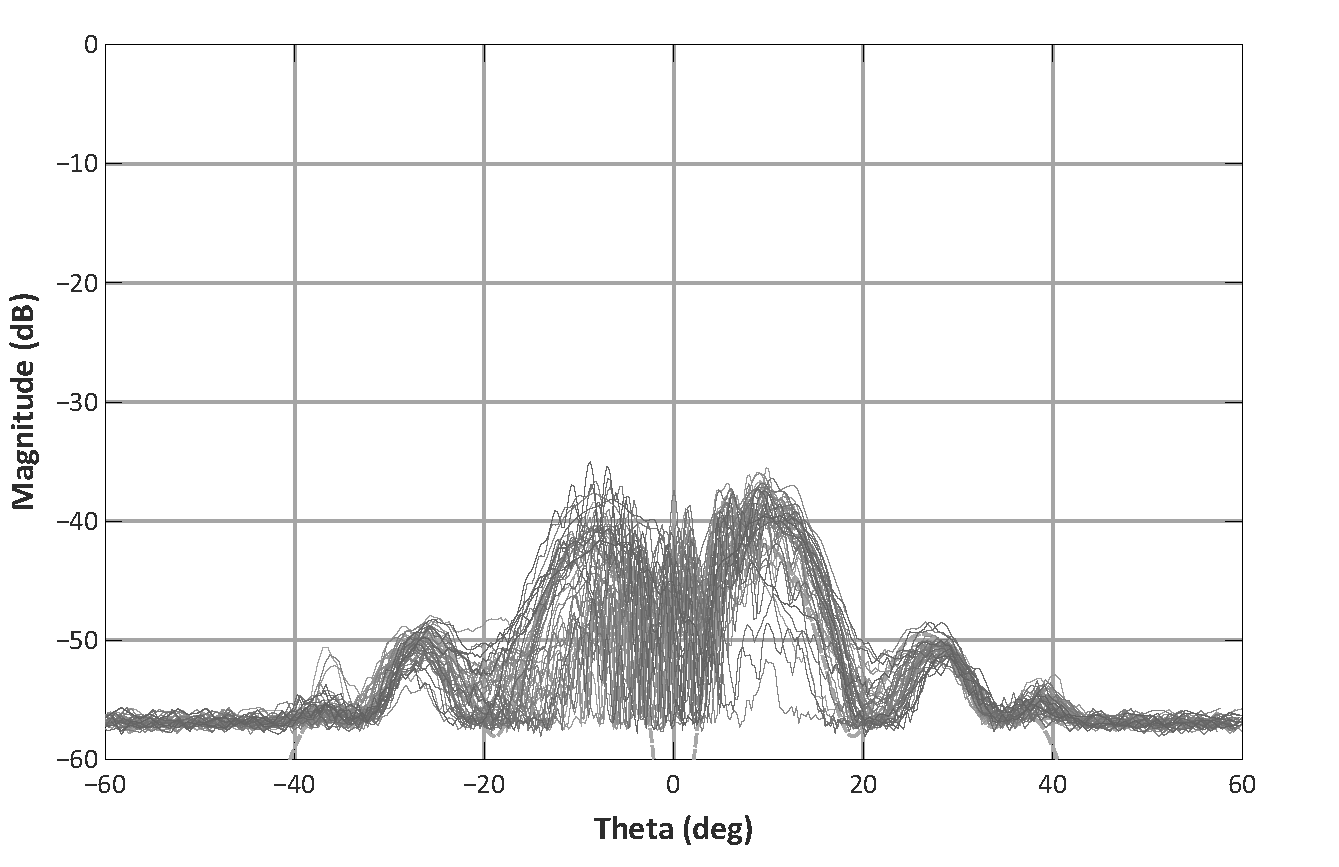}
   \end{tabular}
   \end{center}
   \caption[Measure] 
   { \label{fig:measure} 
Measured radiation patterns of all 49 feedhorns at 43 GHz: normalized co--polar 45$^{\circ}$ plane (left) and cross--polar 45$^{\circ}$ plane (right).}
   \end{figure}

We have measured the feedhorn radiation patterns at six frequencies in the operative bandwidth. Fig.~\ref{fig:measure} shows one cut of the measured radiation patterns for all the 49 feedhorns, and they are highly consistent with one another and in agreement with the simulations within fractions of dB up to the first sidelobe. The directivity increases by $\sim 2$~dB with frequency (from 22.7~dBi at 38 GHz to 24.9 dBi at 48 GHz). Cross-polarization and return loss levels are better than -37 dB and 40 dB, respectively, on the whole bandwidth. 
Measurements have been repeated after a cool-down cycle in liquid Nitrogen, leading to no measurable variations in the performance.

Downstream of each feedhorn, a polarizer converts the linear polarizations into circular polarizations that are, then, routed to two different rectangular waveguides, providing the mechanical mating with the polarimeter module. To this end, the OMT is rotated by 45 deg with respect to the polarizer principal-polarization basis of the polarizer, so that the overall polarizer-OMT pair works like a septum polarizer. The polarizer and OMT designs have been carried out in order to achieve high electrical performance, while providing effective. solutions for large polarimeter arrays, i.e. medium-scale production at affordable cost. The polarizers are based on a split-block layout that guarantees a high mechanical mounting accuracy. More details about the polarizers can be found in Ref.~\citenum{polarizer}. The STRIP turnstile-junction orthomode transducer is based on the layout reported in Ref.~\citenum{omt}, and have been manufactured using the platelet technique with 1 mm-thick layers.

\begin{figure}
   \begin{center}
   \begin{tabular}{c c}
   \includegraphics[width=0.45\textwidth]{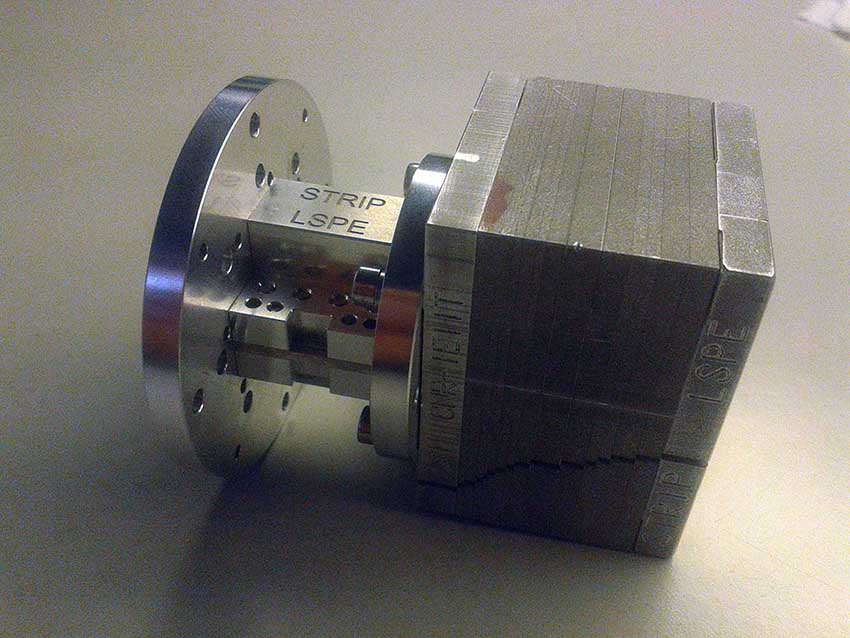} &
   \includegraphics[width=0.45\textwidth]{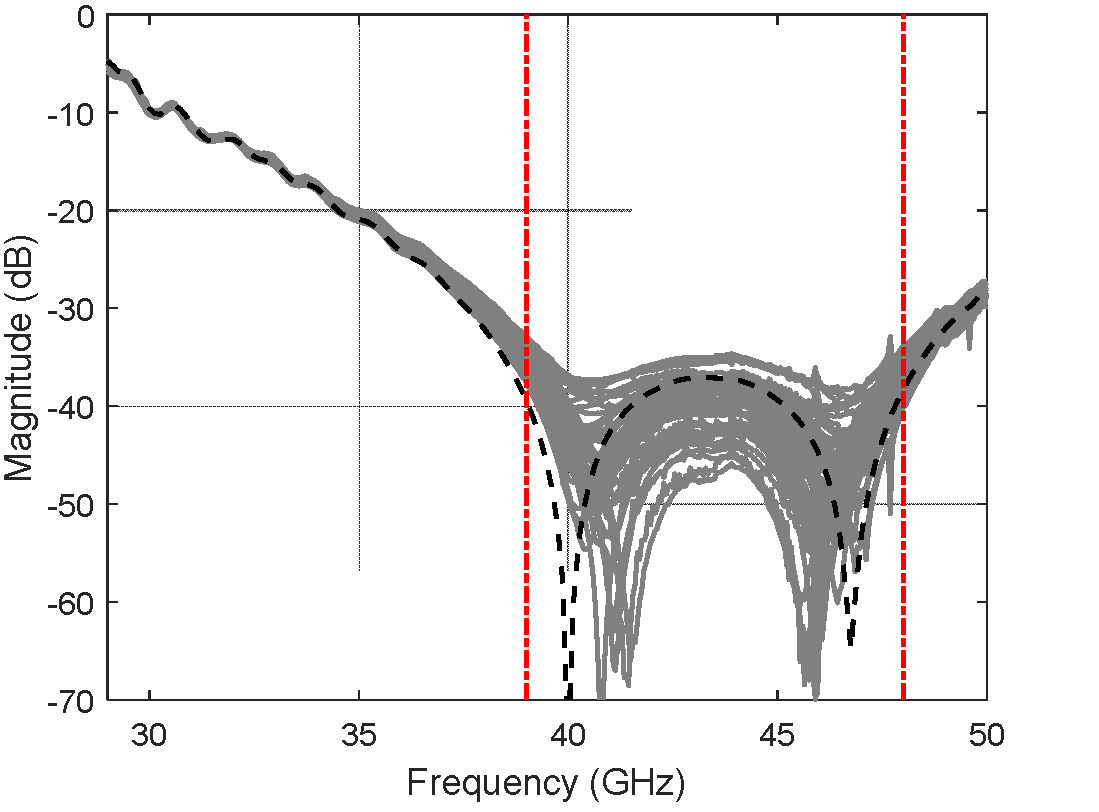} 
   \end{tabular}
   \end{center}
   \caption[Pol_OMT] 
   { \label{fig:pol_omt} 
{\it Left}: Picture of the polarizer and the orthomode transducer assembled together. {\it Right}: Measured levels of cross-polarization of the 49 Q-band sub-assemblies consisting of polarizers and OMTs\cite{polarizer}. Vertical red lines indicate the receivers nominal bandwidth limits.}
   \end{figure}

The measured RF characteristics of the polarizers and OMTs show a remarkable agreement with the predictions. The forty-nine polarizer units present a return loss higher than 38~dB for both the circular polarizations. The measured insertion loss is lower than 0.07~dB with an equalization better than 0.04 dB. The phase shift between the principal polarizations of the polarizers is $90\pm 1.8$ degrees, yielding a cross-polarization transmission between the right-hand and left-hand circular polarizations lower than -35~dB. The forty-nine STRIP OMTs are characterized by a return loss better than 22~dB, isolation between rectangular ports greater than 50~dB and a cross-polarization lower than -50~dB. We made also measurement on the polarizer and OMT sub-assemblies, to check the compliance with the unit tests, characterizing the rejection coefficients, isolation, insertion losses and cross-polarization (see Fig.~\ref{fig:pol_omt}, right panel). A picture of the polarizer/OMT assembly is shown in see Fig.~\ref{fig:pol_omt}, left panel.

The orthogonal components split by the orthomode transducer propagate through a polarimeter module, which amplifies, correlates and detects the signal, enabling the detection of the Q and U Stokes parameters.

 \begin{figure}
   \begin{center}
   \includegraphics[width=0.45\textwidth]{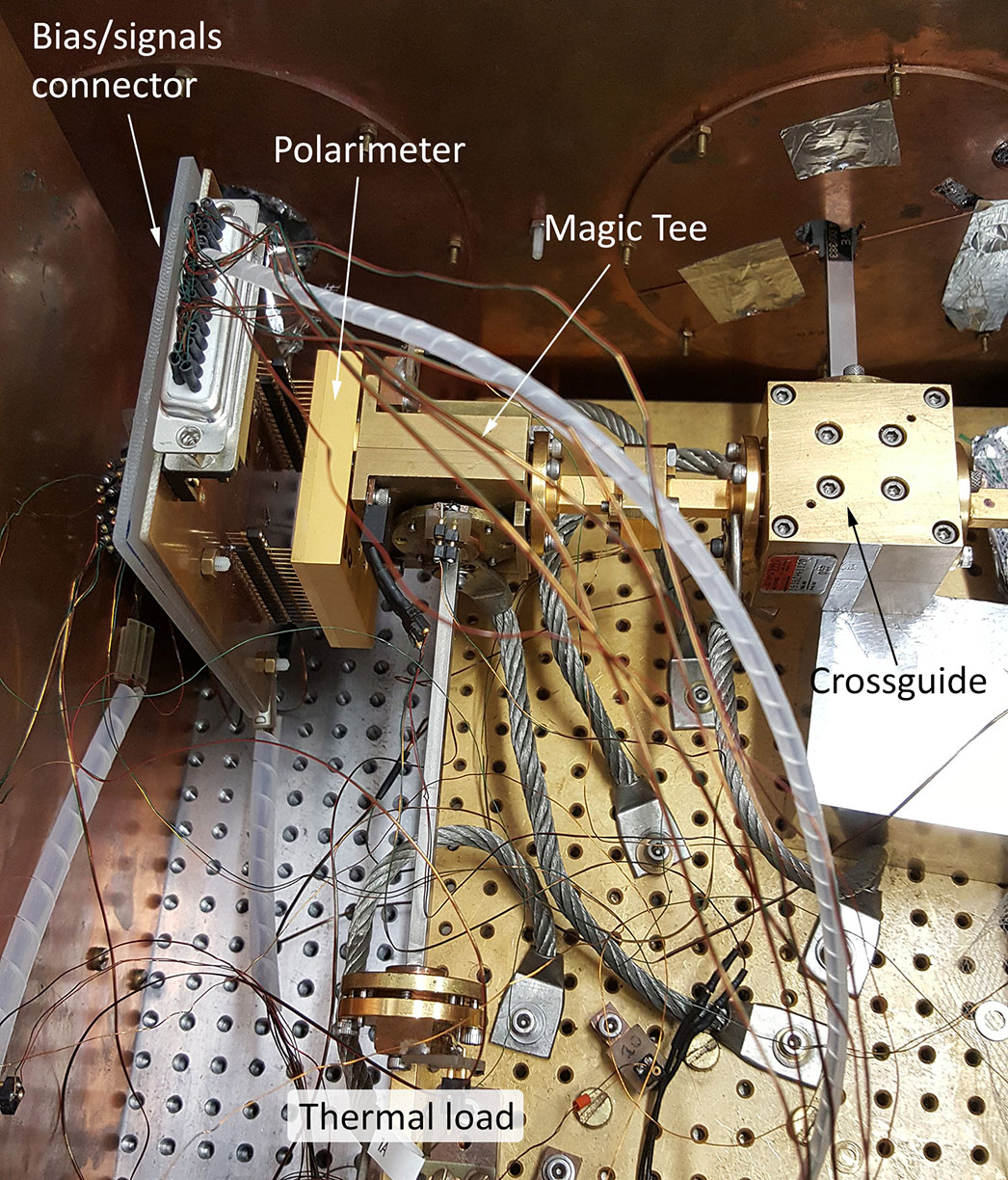}
   \end{center}
   \caption[MMIC_unit_test] 
   { \label{fig:MMIC_unit_test} 
Internal view of the cryo-chamber where STRIP polarimeters have been characterized. On top-left, a Q-band polarimeter is mounted to the connectors plate, while waveguide inputs are fed by a magic-Tee combiner. The cross-guide component couples a tone produced by an external swept source generator to measure the polarimeter bandpass response.}
   \end{figure}

The design of the STRIP polarimeters is based on cryogenic High Electron Mobility Transistor (HEMT) low noise amplifiers and on high-performance waveguide components cooled to 20~K and integrated in Monolithic Microwave Integrated Circuits (MMIC). 

The design is the same as the one proposed for the QUIET experiment\cite{quiet_pol}. In particular, some of the polarimeters in the focal plane are the same devices used in the QUIET instrument, while others have been produced following the same design.

As shown in the schematic configuration in Fig.~\ref{fig:config}, the correlation unit receives the circular polarization coming from the orthomode transducer and each input is independently amplified and phase shifted. One phase-switch shifts the phase between 0$^{\circ}$ and 180$^{\circ}$ at 4~kHz, while the other works at 100~Hz. The two signals are combined in a 180$^{\circ}$ hybrid coupler. Half of each output is bandpass filtered and rectified by a pair of detector diodes, while the other half passes through a 90$^{\circ}$ hybrid coupler. A second pair of bandpass filters and detector diodes measures the power coming out from the coupler. 

The benefit of this design is that we obtain the Q and U Stokes parameters directly from each pair of diodes.
Synchronous demodulation of the 4 kHz phase switching suppresses low-frequency atmospheric fluctuations as well as 1/f noise from the amplifiers and detector diodes. 
Subsequent demodulation of the 100 Hz phase switching removes spurious instrumental polarization generated by unequal transmission coefficients in the phase-switch circuits. Averaging the output of each diode rather than demodulating it results in a measurement of the total intensity.

 \begin{figure}
   \begin{center}
   \begin{tabular}{c c}
   \includegraphics[width=0.48\textwidth]{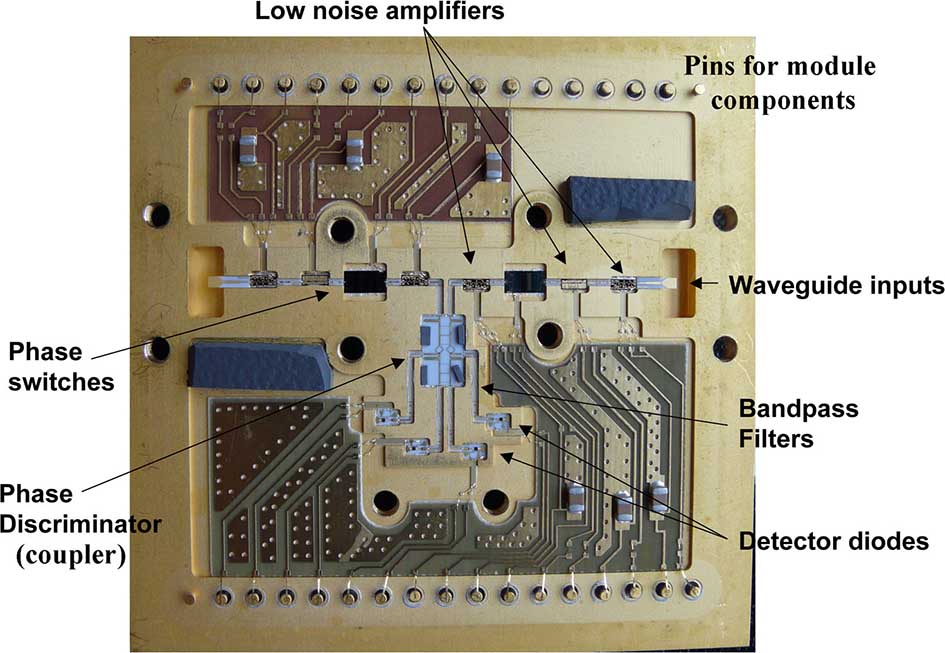} &
   \includegraphics[width=0.49\textwidth]{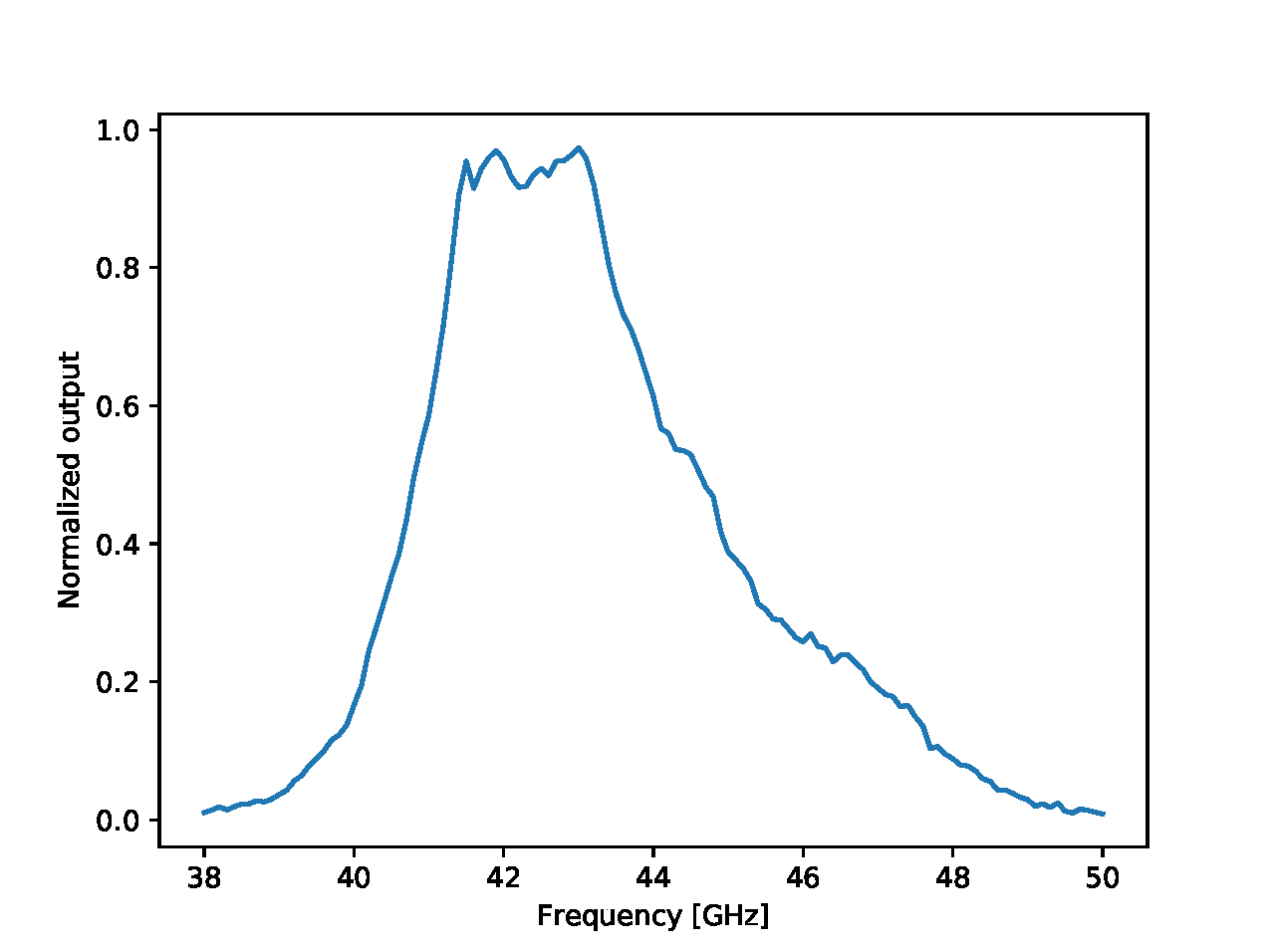} 
   \end{tabular}
   \end{center}
   \caption[Polarimeter] 
   { \label{fig:polarimeter} 
{\it Left}: Internal view of a 43 GHz polarimeter module\cite{quiet_pol}. The correlator components are assembled in a single Monolithic Microwave Integrated Circuit (MMIC). {\it Right}: Measured bandpass for a STRIP polarimeter. The shaded line gives the maximum variation of the bandpass, measured as the output of each diode of the receiver.}
   \end{figure} 
   
Polarimeters have been characterized with functional and performance tests, both at room temperature and in cryogenic conditions to evaluate the polarimeters functionality and performance. Fig.~\ref{fig:MMIC_unit_test} shows a Q-band polarimeter as mounted in the cryo-chamber during unit test measurements.

Functionality tests include the bias response of the polarimeter HEMTs, phase switches and diode detectors as well as the characterization of the raw voltage output in reference bias conditions.
Performance tests include measurements of the noise temperature, of the receiver bandpass and of its noise properties (white and 1/f noise).
The average bandpass is 7.3 GHz, centered at 43.3 GHz. An example of a measured bandpass is shown in Fig.~\ref{fig:polarimeter}.

\subsubsection{The W-band Radiometer Chain}
\begin{figure}
   \begin{center}
   \begin{tabular}{c c}
   \includegraphics[width=0.40\textwidth]{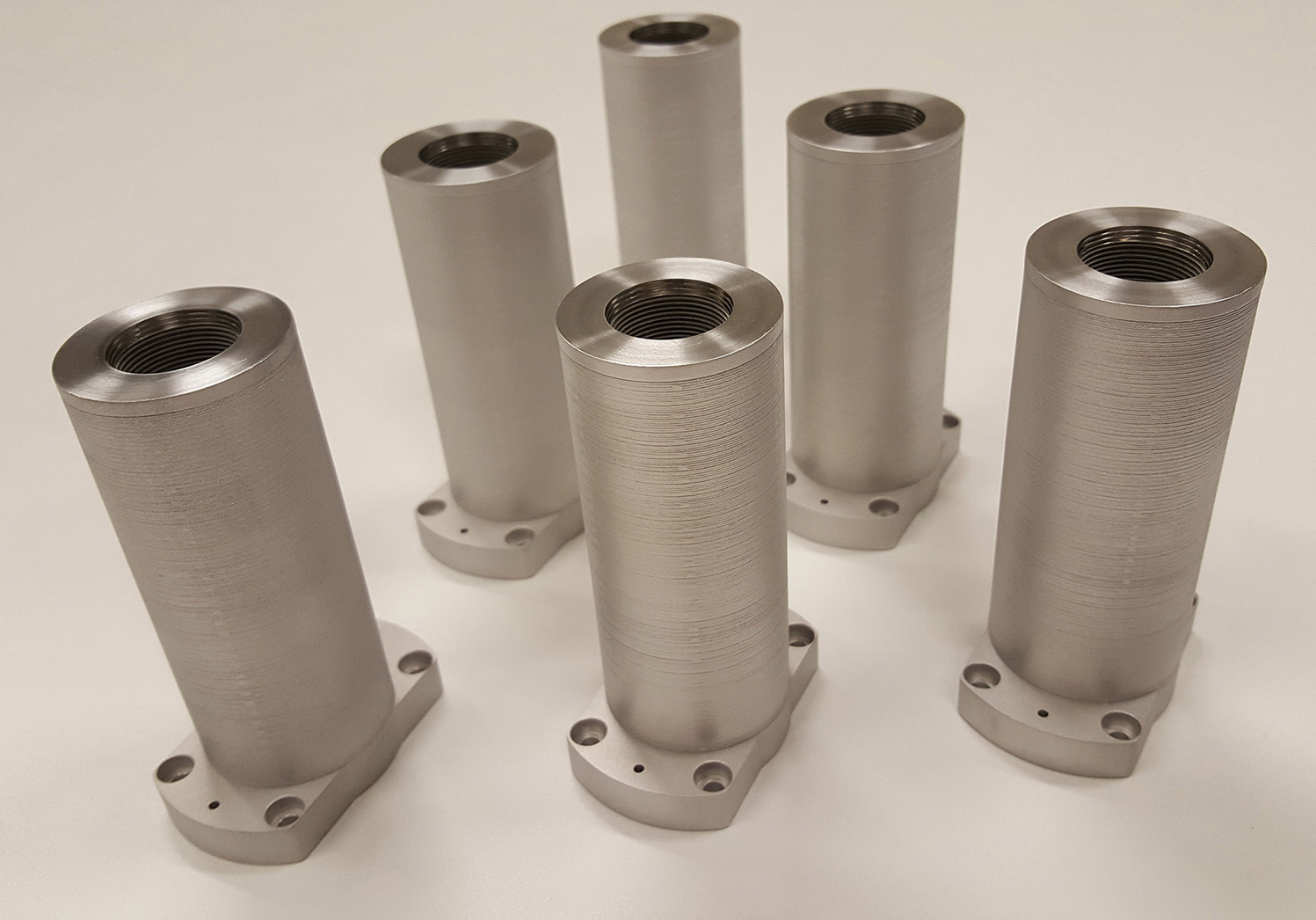} &
   \includegraphics[width=0.55\textwidth]{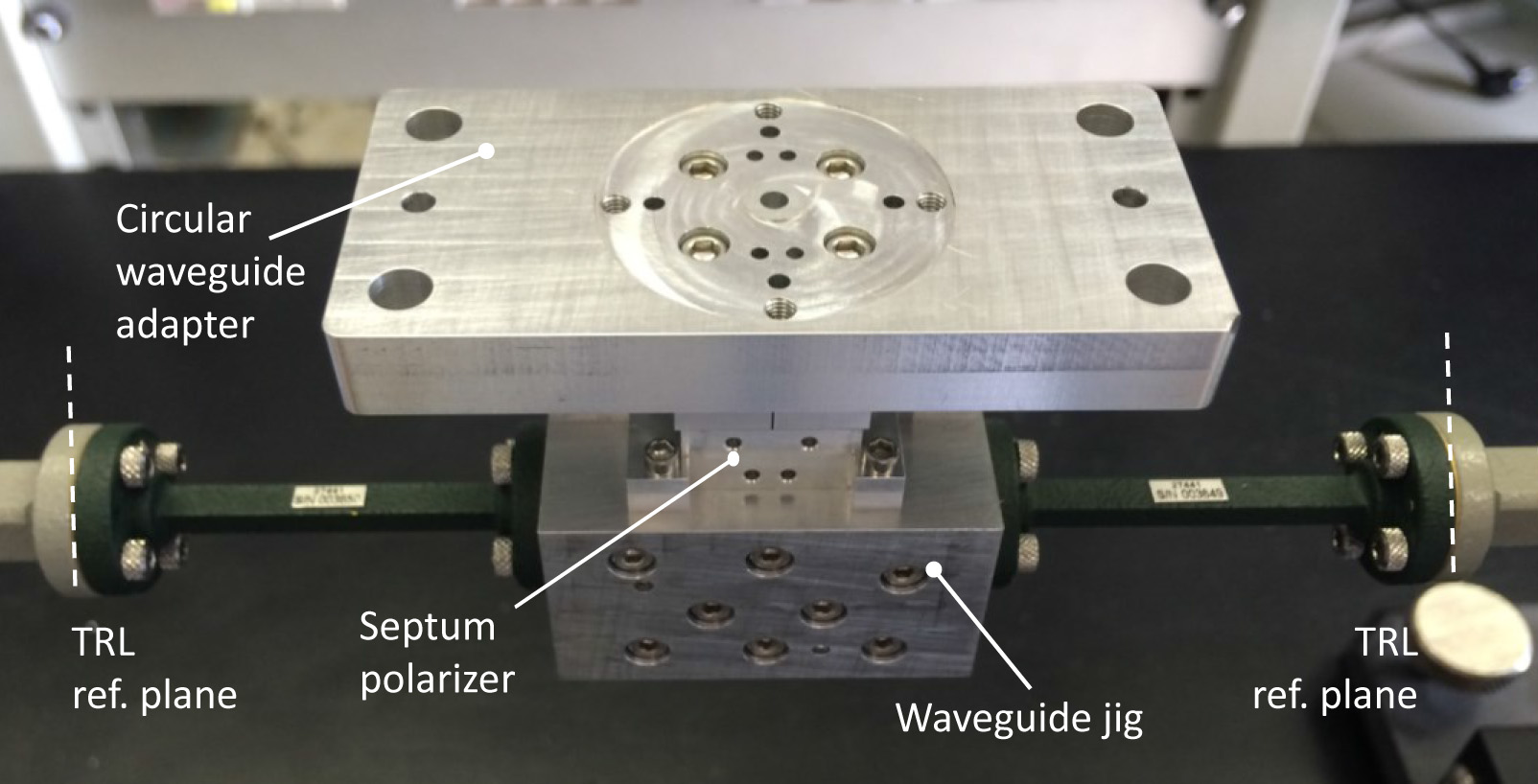}
   \end{tabular}
   \end{center}
   \caption[Wband_HW] 
   { \label{fig:Wband_HW} 
W-band passive components. {\it Left}: The six platelet feedhorns. {\it Right}: A septum polarizer during unit test characterization.}
   \end{figure}

The STRIP W-band channel will be used as an atmospheric monitor, especially to track the amount of water vapor, which is the primary cause of atmospheric opacity. 
Circular corrugated feedhorns have been chosen as the best solution for the W-band channels as well. 
Each feedhorn has been built using the platelet technique, where the layers have been realized with a chemical etching technique. Left panel of Fig.~\ref{fig:Wband_HW} shows the six feedhorns array before starting their characterization in the anechoic chamber. At the time of writing this manuscript, we are finalizing the radiation pattern and return loss measurements.

The radiation from each feedhorn enters a septum polarizer\cite{septum} which separates the incoming signal into left- and right-circularly polarized components and it is designed to allow the widest range of frequencies. Then, the two components are separated into two waveguide ports which mate to a correlation polarimeter module. The design of the W-band sub-assembly requires also an adapter flange to mate the QUIET original septum polarizer (see Fig.~\ref{fig:Wband_HW}, right panel), which acts like the Q-band sub-assembly of polarizer and OMT, to the W-band feedhorn.

Since STRIP uses six W-band polarimeters of the QUIET experiment, we refer the reader to the technical description in Ref.~\citenum{QUIET_polarimeters}.

\subsection{Calibration Sub-systems: Near-Field Calibrator and Star Tracker} 
\subsubsection{Near-Field calibrator} 
To meet specific needs in terms of instrumental requirements, such as relative gain stability and telescope pointing accuracy. In particular, a ``near-filed'' or internal calibrator (NFC), will provide a stable relative calibration signal throughout the duration of the observations. The calibration signal will be produced by a couple of oscillators with electrically tunable carrier frequency, a varactor tunable oscillator in the Q-band and a Gunn oscillator in the W-band. Each tone will be fed into a horn, 
placed in the region between the two reflectors of the telescope and directly illuminating the focal plane array.

Given the high power level of the oscillators (of order of 10-12 dBm), the sources will be suitably isolated and attenuated to avoid receivers saturation. Internal heaters will provide temperature stability to the sources. Susceptibility to temperature oscillations is of order of 0.01 dB/$\degree$C, with an expected long term temperature stability of 0.05 dB/day. Sources will be biased by STRIP electronics and will be switched ON/OFF during operations, according to the calibration strategy.

\subsubsection{Star Tracker} 
The telescope pointing accuracy requirement will be addressed by another ancillary system, the Star Tracker. Mounted on the baffle of the telescope, the star tracking system will allow to accurately reconstruct the transformation matrix which turns the reference frame of the telescope (with respect to the motor encoders) into the celestial sky frame. This will be achieved by observing the apparent position of an adequate number of bright stars in the optical/infrared band. Star tracker will include a high-speed guiding camera equipped with an electronic global shutter with exposure times as short as 1/100 sec, a diagonal of about 1400 pixels, and with a download capability of 10 frames per second at least. The camera will be coupled to a refractor telescope, resulting in a field of view about $3\degree-5\degree$, good enough to properly detect the positions of at least three stars.

A set of inclinometers mounted on the basement of the telescope will be used to monitor the settlement of concrete in the months after the building of the telescope and tilts due to seismic waves. It is therefore not needed to calibrate them with respect to some absolute reference frame, as their purpose is to provide a relative measurement with respect to a zero point (e.g. when the alignment of the telescope is complete). The inclinometers mounted on the telescope will be calibrated by observing some star at a known zenithal distance and comparing the measurement of the angle provided by the encoders with those provided by the inclinometers. Several stars picked at different azimuth shall be used in the measurement, in order to properly verify any unwanted tilt of the telescope main axis.

\subsection{Cryostat} 
\begin{figure}
   \begin{center}
   \includegraphics[width=1.0\textwidth]{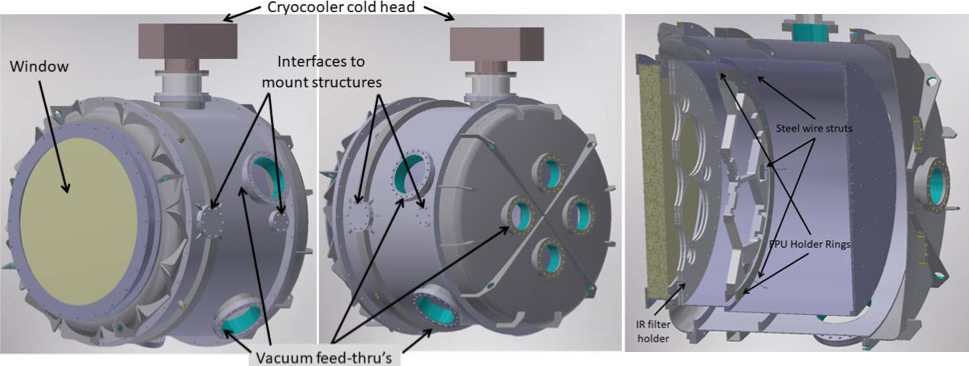}
   \end{center}
   \caption[cryostat]
   { \label{fig:cryostat} 
Views of the STRIP cryostat. {\it Left and center}: Front and rear views of the CAD project, including the Ultra-high Molecular Weight Polyethylene window and vacuum feed-thru's. {\it Right}: Steel wire struts for the FPU frame. The wires support two circular interfaces that holds the FPU frame in order to minimize the mechanical stresses on the frame.}
   \end{figure}

The STRIP cryostat (see Fig.~\ref{fig:cryostat}) is designed to ensure the thermal environment required to operate the instrument at the observing site, hence it should cool the focal plane unit down to 20 K. 
Moreover, it should operate with an external temperature that typically ranges between 0$^{\circ}$ and $24^{\circ}$ and an average pressure of 760 mbar.

Both the detectors and the radiative shield (at the operating temperatures of 20 K and 100 K, respectively) are cooled by means of a two-stage mechanical cryocooler. The first stage, in the 70-110 K range, cools the radiative shield and intercepts the parasitic heat leaks due to radiation, harness or mechanical struts. The second stage, the coldest one, provides the 20 K reference for the polarimeters array. The vacuum pressure required by the cryostat is $P \le 10^{-4}$ mbar.
   
One of the main issues of the cryostat design optimization is the minimization of the parasitic heat leaks to colder stages. All main heat transfer paths through conduction and radiation have been evaluated and reduced by optimizing material selection and the mechanical configuration.

The focal plane and the radiative shield are mechanically supported with a 2~mm thick steel wire, ensuring both mechanical stiffness and thermal insulation. The structural supports are made of low thermal conductivity materials with optimized dimensions and thickness. In order to minimize the loads, two circular interfaces are actually supported by the pre-tensioned wires and hold the focal plane frame in place, maintaining its deflection with respect to the window axis within 1~mm.
The focal plane is thermally connected to the 20~K cooling stage by means of oxygen-free copper thermal straps linked to a copper flange integrated on the cooler cold tip.
The large number of electrical connections (around 1500 wires) requires a careful cryoharness thermal design, since it is one of the major contributions to parasitic loads. For this reason, the radiative shield is designed to allocate contact areas to thermalize the harness and intercept the leaks on the first stage of the cooler, minimizing the load on the 20K stage.

The cryostat size is driven by the FPU diameter, the window opening on the outer shell and the IR blocking filter on the 100 K shield. It is designed to minimize the envelope and mass of the system, in order to avoid over stress on the telescope mount structure. For these reasons the outer shell and the radiative shield are built using stainless steel and aluminum respectively. Both the shield and outer shell internal surface have a low emissivity coating. The cryostat window has been designed using Ultra-high Molecular Weight Polyethylene (UHMWP).

The cryostat mechanical configuration is strong enough to withstand the atmospheric pressure and the loads originated by supporting the internal units and the whole system in its position during the observations.

\subsection{Electronics} 
\begin{figure}
   \begin{center}
   \begin{tabular}{c c}
   \includegraphics[width=0.35\textwidth]{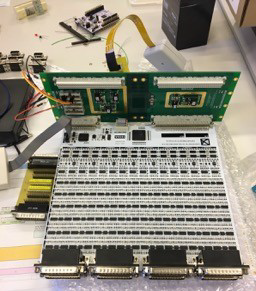} &
   \includegraphics[width=0.45\textwidth]{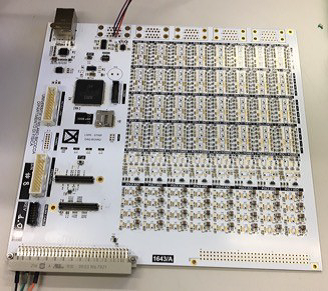}
   \end{tabular}
   \end{center}
   \caption[Board] 
   { \label{fig:board} 
Twin boards test phase. {\it Left}: Eight polarimeters bias board attached to the power supply back-plane. {\it Right}: Eight polarimeters preamp and readout board.}
   \end{figure}

Seven twin boards are necessary to drive and acquire 49 Q-band and 6 W-band polarimeters. Figure \ref{fig:board} shows a prototype of the boards during the functionality and performance verification. Each unit is dedicated to eight polarimeters and is in charge of the biasing (48 InP HEMT LNAs, 16 InP phase shifters and 32 GaAs detectors) and data acquisition. The acquisition board has 32 high sensitivity low noise preamplifiers, with bias and offset control, and 18 bit ADCs at 1.6 Msps and it is equipped with very powerful logic (FPGA) to perform pre-analysis (high frequency demodulation) on the acquired data and locally store the data. A programmable logic controller (PLC) is in charge of data IO, getting telecommands from the control computer and organizing data packets for eternal storage. The units transfer the data to the CPU Unit via Ethernet LAN.
\subsection{Software} 
The STRIP control software is in charge of communications (telecommands and telemetry through the dedicated ethernet line) from/to the instrument and telescope mount. It is based on a web portal that provides a simple and effective user interface to the operator, allowing (1) to monitor the instrument, browsing through live and past data; (2) send commands to the instrument boards and the cryogenic unit; (3) monitor the status of the telescope mount and view/modify the observation plans.

\begin{figure}
   \begin{center}
   \includegraphics[width=0.9\textwidth]{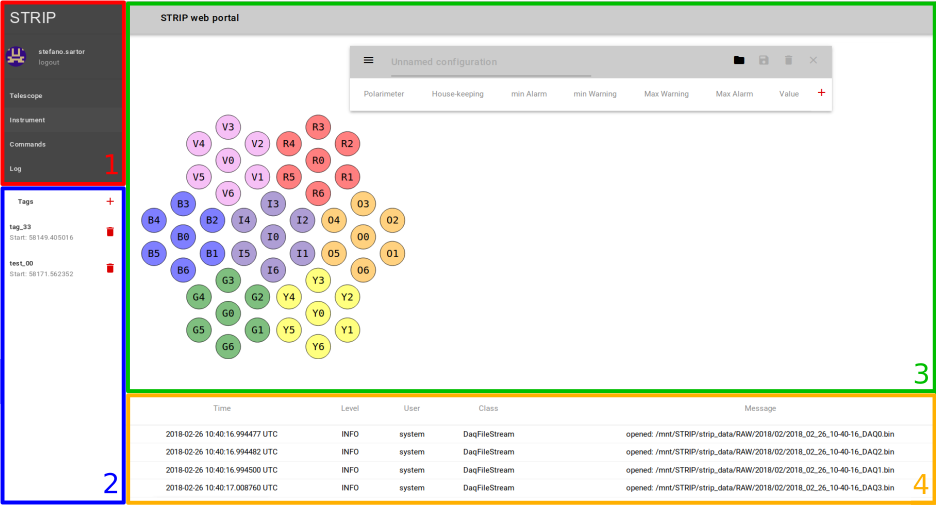}
   \end{center}
   \caption[sw] 
   { \label{fig:sw}
   STRIP web portal main structure.}
   \end{figure}

The Web portal is divided in 4 main sections:
\begin{itemize}
\item \textbf{navigation bar}: allows the user to login and to navigate through the pages of the portal, as in panel 1 of Fig.~\ref{fig:sw};
\item \textbf{tag list}: shows the current active tags, with the start timestamp expressed in modified Julian date, as in panel 2 of Fig.~\ref{fig:sw};
\item \textbf{main section}: displays the information relative to the link highlighted in the navigation bar (Telescope, Instrument, Commands and Log), as in panel 3 of Fig.~\ref{fig:sw};
\item \textbf{log table}: allows the user to browse, filter and query log events, as in panel 4 of Fig.~\ref{fig:sw}.
\end{itemize}

All the data displayed in the pages of the STRIP portal are fetched via REST requests or websockets. A superset of those APIs are available for user-scripting (in any programming language, as long as it supports REST and websocket protocols). The websocket API is used in the STRIP web portal to provide all the asynchronous events and data streaming.

\section{CONCLUSIONS} 
\label{sec:conclusion}

At present, most of the STRIP hardware has been developed and tested. Q-band and W-band sub-systems, including passive components (feedhorns, polarizers, OMT's) and HEMT-based coherent receivers, have been successfully characterized at unit-level. Integration and system-level testing of the focal plane arrays inside the instrument cryostat is planned to be performed within the next months, starting in July. Electronics and ancillary systems, such as the internal calibrator and the star tracker, are being finalized and verified as well. Integration and verifications of the telescope on its mount are proceeding, leading STRIP to be fully integrated at the observation site within the end of 2018.

After the on-site instrument verification and calibration, STRIP will be operative for a 2-years campaign of observation of the CMB polarization.

In this paper, we discussed the status and the latest developments of STRIP, the ground-based instrument of the “Large Scale Polarization Explorer” experiment. By combining ground-based (STRIP) and balloon-borne (SWIPE) polarization measurements of the microwave sky on large angular scales, LSPE attempts a detection of the so-called ``B-modes'' of the Cosmic Microwave Background polarization, improving the limit on the tensor-to-scalar ratio down to $r$ = 0.03 at 99.7\% C.L. Moreover, a second target is to produce wide maps of foreground polarization generated in our Galaxy by synchrotron emission and interstellar dust emission. These will be important to map Galactic magnetic fields and to study the properties of ionized gas and of diffuse interstellar dust in our Galaxy.

Originally designed to observe from the LSPE stratospheric balloon, STRIP has been recently redesigned as a ground-based telescope that will operate from the ``Observatorio del Teide'' in Tenerife. In its new configuration, STRIP will observe approximately 20-25\% of the Northern sky for at least one year, by exploiting a 1.5 m fully rotating crossed Dragone telescope coupled to an array of forty-nine HEMT-based coherent polarimeters, operating at 43 GHz and with an angular resolution of 20'. A second frequency channel with six-elements at 95 GHz will be exploited as an atmospheric monitor.


\acknowledgments     
 
This work has been carried out in the context of the Italian Space Agency (ASI) programme ``Large Scale Polarization Explorer (LSPE)''. We gratefully acknowledge support from ASI through contract I-022-11-0 ``LSPE'' and from the Istituto Nazionale di Fisica Nucleare (INFN).




\bibliographystyle{spiebib}   

\end{document}